\documentclass[lineno]{ankit_paper}

\usepackage{natbib}
\usepackage{xcolor}
\usepackage{float}

\lefttitle{A. Bajpai and J. Gopalan}
\righttitle{Turbulent Spots in Hypersonic Boundary Layers}

\title{Turbulent spots in hypersonic transitional planar and axisymmetric boundary layers}

\author{Ankit Bajpai\aff{1} and Jagadeesh Gopalan\aff{1}}

\affiliation{\aff{1}Department of Aerospace Engineering, Indian Institute of Science, Bengaluru, Karnataka, INDIA}

\corresau{Ankit Bajpai, \url{ankitbajpai23@gmail.com}}

\begin{document}
\maketitle

\begin{abstract}
Experiments were conducted to investigate characteristics of turbulent spots formed in transitional boundary layers developed over a flat plate and an axisymmetric cone placed in similar hypersonic freestream environment of Mach number 5.85. The freestream Reynolds number in the present work varied between $3.0-6.0\times10^6$/m. Heat transfer measurement along the surface of both the test models was used to ascertain the state of boundary layer and to calculate the intermittency associated with transitional boundary layer. Turbulent spots generated in the transitional boundary layer were characterized in terms of their leading-trailing edge velocities, their streamwise length scales and their generation rates on both the test models. Leading edge of the turbulent spots developed over both the test models were found to be convecting at a speed equivalent to 90\% of the boundary layer edge speed. The trailing edge of the spots developed on a planar boundary layer traversed at a lower speed than its axisymmetric counterpart. Streamwise length scales of a turbulent spot developed in a planar boundary layer grew at a higher rate when compared with axisymmetric boundary layer. Turbulent spot generation rates for both planar and axisymmetric boundary layers was found to be in the range of $1000000-3000000$ spots/m/s. 
\end{abstract}

\begin{keywords}
Intermittency, Turbulent Spots, Hypersonic Flow, Compressible boundary layers
\end{keywords}


\section{Introduction}
\label{sec:headings}

The transition from laminar boundary layer to turbulent boundary layer occurs through spatially sporadic onset of turbulence. These isolated regions of turbulence superimposed over laminar boundary layer are called `turbulent spots'. These spots are convected downstream from the location of their inception during which they grow in size and merge to result in a fully developed turbulent boundary layer. In case of high speed flows, the area of a surface over which the generation and growth of these turbulent spots occur can be substantially larger than the low speed flow regime. Hence the extent of this intermittent region of the boundary layer becomes an important input in the design of air breathing hypersonic vehicles wherein the state of the boundary layer can affect the performance of the vehicle.\\
The observations and analysis by Emmons \citeyearpar{Emmons51} laid the foundations of a theory of transition in boundary layer. The transition from laminar to turbulent boundary layer occurs by the turbulence spread from the islands of turbulence in the pool of laminar boundary layer. The theory postulated that the transitional boundary layer was of considerable length and acted as a bridge between the two stables states of boundary layer namely laminar and turbulent. Narasimha \citeyearpar{Nara57} and several others further expanded the work of Emmons \citeyearpar{Emmons51} which included a modified definition of intermittency function by using Dirac delta function to approximate source rate density and experiments in incompressible, subsonic, supersonic flow regime. A review of earlier works on turbulent spots in subsonic and low supersonic flow can be found in Narasimha \citeyearpar{Nara85}. \\
A general outline and recommendations for experimental measurement of intermittency was given by Hedley and Keffer \citeyearpar{Hedley74} and Narasimha \citeyearpar{Nara85}. Experimental measurements on intermittent nature of compressible high speed transitional boundary layers have been usually carried out by using thermal sensors like thin film sensors or thermocouples. Intermittency measurements in a transitional boundary layer developed on a $5^{\circ}$ circular cone under hypersonic freestream conditions was reported by Owen and Horstman \citeyearpar{Owen72} and it was found that the intermittency distribution on the cone closely followed the incompressible distribution. The first investigation on characterizing turbulent spots developed on a flat plate transitional boundary layer was carried out by Clark, Jones and LaGraff \citeyearpar{Clark94}. The study was carried out for a range of freestream Mach numbers 0.24-1.86 on both zero pressure gradient and favorable pressure gradient boundary layers. The study concluded that the presence of a favorable pressure gradient in the boundary layer lead to an increase in spot trailing edge velocity while spot leading edge velocity remained unaffected by the pressure gradient. Heat transfer measurements to identify turbulent spots were carried out in hypersonic transitional boundary layers developed over a cylinder with a spherical nose by Fiala \textit{et al.} \citeyearpar{Fiala06}. The investigation by Fiala \textit{et al.} \citeyearpar{Fiala06} measured turbulent spot propagation speeds and found a internal cell structure within the turbulent spot. It also remarked on the many similarities between spot dataset obtained at hypersonic flow speeds and the studies conducted at comparatively lower flow speeds. Jewell \textit{et al.} \citeyearpar{Jewell12} reported several findings on turbulent spots observed in a transitional boundary layer developed over a $5^{\circ}$ half angle cone. The experiments were conducted in T5 hypervelocity reflected shock tunnel. Turbulent spot leading and trailing edge velocities were measured and compared with other previously conducted works discussed over here as well. Mee and Tanguy \citeyearpar{mee15} measured spot initiation rates in transitional boundary layers of a flat plate placed in hypersonic freestream of a reflected shock tunnel (T4 Stalker tube). This is the only work, in open literature to the best of authors knowledge, wherein spot initiation rates were calculated and a correlation for the same was proposed. Lunte and Sch\"{u}lien \citeyearpar{lunte25} measured intermittency in the transitional boundary layer developed on a flat plate placed in a hypersonic freestream. The intermittency measurements were conducted using three different techniques, namely heat flux density measurement, density fluctuations measurement and pressure fluctuations measurement. The intermittency measured by the three techniques were compared with each other as well as the universal intermittency and the match was found to be mutually consistent barring the intermittency distribution obtained from pressure fluctuations measurement. This deviation was attributed to intrusive measurement involved in pressure fluctuations measurements. The turbulent spots in hypersonic transitional boundary layers have also been studied numerically using Direct Numerical Simulations (Krishnan and Sandham \citeyearpar{krishnan06}, Hader and Fasel \citeyearpar{hader19}). The numerical simulations revealed turbulent spots to have an upstream-pointed arrowhead shape with a leading-edge overhang, followed by a turbulent core and a calmed region at the rear interface.\\
As evident from the brief literature review discussed in previous paragraph, the intermittent nature of hypersonic transitional boundary layer has been probed by many research groups but the quantum of work is less exhaustive when compared to studies on boundary layer intermittency conducted for subsonic and low supersonic freestream environment. Earlier work in this context revealed some important outcomes pertaining to a particular test model and ground test facility freestream environment. A comparison of turbulent spot characteristics for hypersonic transitional boundary layers developed over different surfaces under similar freestream conditions is seldom reported in the literature. In the case for hypersonic boundary layers, dataset for spot generation rate of turbulent spots is scarce. These observations led to the present work which involves characterization of turbulent spots developed over two different generic test bodies kept under same freestream conditions. Flat plate with sharp leading edge and an axisymmetric cone with a sharp tip were chosen as two test models to compare turbulent spot characteristics on a planar and a conical boundary layer. The test models were placed in the test section such that the boundary layer edge Mach number and Reynolds number for both the cases were comparable. Spot generation rates were also calculated from the experimental data thus obtained.

\section{Experimental Details}
\subsection{Hypersonic Shock Tunnel 4 (HST4) Facility}
\begin{figure}[h]
  \centerline{\includegraphics[width=0.75\textwidth]{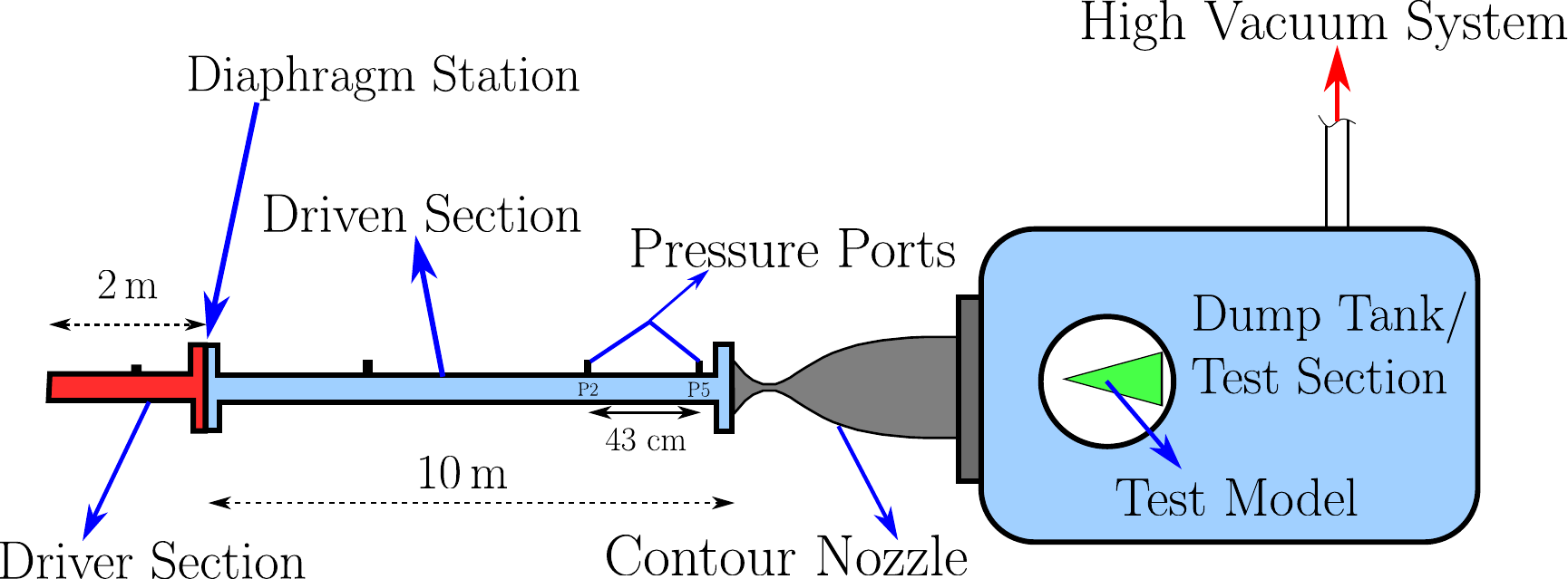}}
  \captionsetup{style=capcenter}
  \caption{Figure depicting schematic of HST4}
\label{fig:hst4_schm}
\end{figure}

The experiments in the present work were conducted in Hypersonic Shock Tunnel 4 (HST4) of Laboratory for Hypersonic and Shockwaves Research (LHSR) at IISc. HST4 is a pressure driven shock tunnel equipped with an axisymmetric contour nozzle capable of generating hypersonic flow in the test section for a useful test duration of 3.5 ms. A schematic of the facility is shown in \hyperref[fig:hst4_schm]{figure \ref{fig:hst4_schm}}. HST4 in the present work is operated to result in a freestream Mach number of $5.8\pm0.1$. The freestream Reynolds number varied in the range of $3.0-6.0\times10^{6}/m$. Further details on the facility can be found in Bajpai and Jagadeesh (\citeyear{bajpai23}).
\subsection{Models, Test Conditions and Instrumentation}
\begin{figure}[h]
  \centerline{\includegraphics[width=0.75\textwidth]{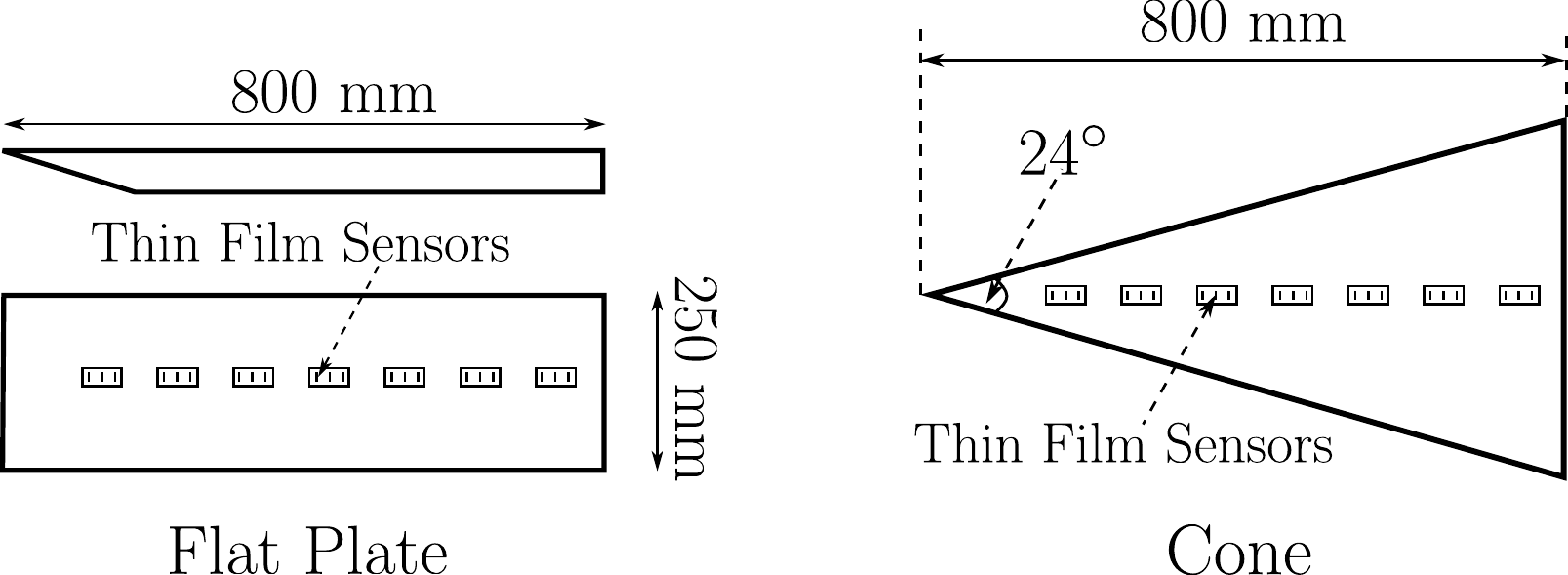}}
  \captionsetup{style=capcenter}
  \caption{Figure depicting schematics of flat plate and cone used in present work.}
\label{fig:model_schm}
\end{figure}

A flat plate with a sharp leading edge and a right angled cone with a sharp tip have been used as the test models in the present work. The length of the flat plate and the axial length of the cone was 800.0 mm. The test models were fabricated with duralumin with stainless steel leading edge and cone tip. The surface roughness of the test models was maintained at N10 ISO grade and the mating surfaces had a tolerance of $50\pm10 \mu m$. Both the leading edge of the flat plate and tip of cone were machined to a sharpness of $80\pm10 \mu m$. The test models were equipped with 20 platinum thin film sensors for wall heat transfer measurements. The heat transfer measurement is used to ascertain the state of the boundary layer at a given spatial and temporal location. The distance between the thin film sensors varied along the length of the test model (from $10.0 mm$ to $70.0mm$) such that the sensors can monitor heat transfer levels in all the three states of the boundary layer namely laminar, transitional and turbulent. The output from heat transfer sensors was acquired at $1.0 MSa/s$ with a NI PXI 6133 DAQ. \hyperref[fig:model_schm]{Figure \ref{fig:model_schm}} shows the schematic of the test models along with the typical placement of thin film sensors used in the present study. 

Similar freestream conditions were maintained over both test models for various cases considered in the present work. The boundary layer edge conditions over both the test models were calculated using inviscid relations. The boundary layer edge conditions over the cone are governed by the semi angle of the cone which is $12^{\circ}$. The flat plate in the present work was kept at an angle of attack of $10^{\circ}\pm0.5^{\circ}$ so that the edge Mach number ($M_{e}$) and edge Reynolds number ($Re_{e}$) over the plate can match with corresponding edge conditions of the cone. \hyperref[tab:fp_edge]{Tables \ref{tab:fp_edge}} and \ref{tab:cone_edge} show the boundary layer edge conditions maintained over flat plate and cone respectively in the experiments. 

\begin{table}
    \begin{center}
\def~{\hphantom{0}}
  \begin{tabular}{ccccc}
      Case & $M_{e}$  & $P_{e}$ (Pa)   &   $h_{o\infty}$ (MJ/kg) & $Re_{e}$ (million/m) \\[3pt]
       CF1 & $4.6\pm0.1$   & ~~$2600\pm50$~~  & ~~$0.95\pm0.02$~~ & ~~$5.0\pm0.1$~~\\
       CF2 & $4.6\pm0.1$   & ~~$3100\pm70$~~  & ~~$0.83\pm0.02$~~ & ~~$7.0\pm0.2$~~\\
       CF3 & $4.6\pm0.1$   & ~~$3550\pm90$~~  & ~~$0.75\pm0.01$~~ & ~~$9.0\pm0.2$~~\\
  \end{tabular}
  \caption{Typical boundary layer edge conditions for flat plate.}
  \label{tab:fp_edge}
 \end{center}
\end{table}

\begin{table}
    \begin{center}
\def~{\hphantom{0}}
  \begin{tabular}{ccccc}
      Case & $M_{e}$  & $P_{e}$ (Pa)   &   $h_{o\infty}$ (MJ/kg) & $Re_{e}$ (million/m) \\[3pt]
       CC1 & $4.6\pm0.1$   & ~~$2400\pm50$~~  & ~~$0.94\pm0.02$~~ & ~~$5.0\pm0.2$~~\\
       CC2 & $4.6\pm0.1$   & ~~$2900\pm70$~~  & ~~$0.82\pm0.02$~~ & ~~$7.0\pm0.2$~~\\
       CC3 & $4.6\pm0.1$   & ~~$3300\pm90$~~  & ~~$0.73\pm0.01$~~ & ~~$9.0\pm0.2$~~\\
  \end{tabular}
  \caption{Typical boundary layer edge conditions for cone. (No transition onset observed for CC1)}
  \label{tab:cone_edge}
 \end{center}
\end{table}

\section{Results and Discussion}
\subsection{Heat Transfer and Intermittency Calculations}
\label{sec:ht_inter_calc}
\begin{figure}[H]
  \centerline{\includegraphics[width=\textwidth]{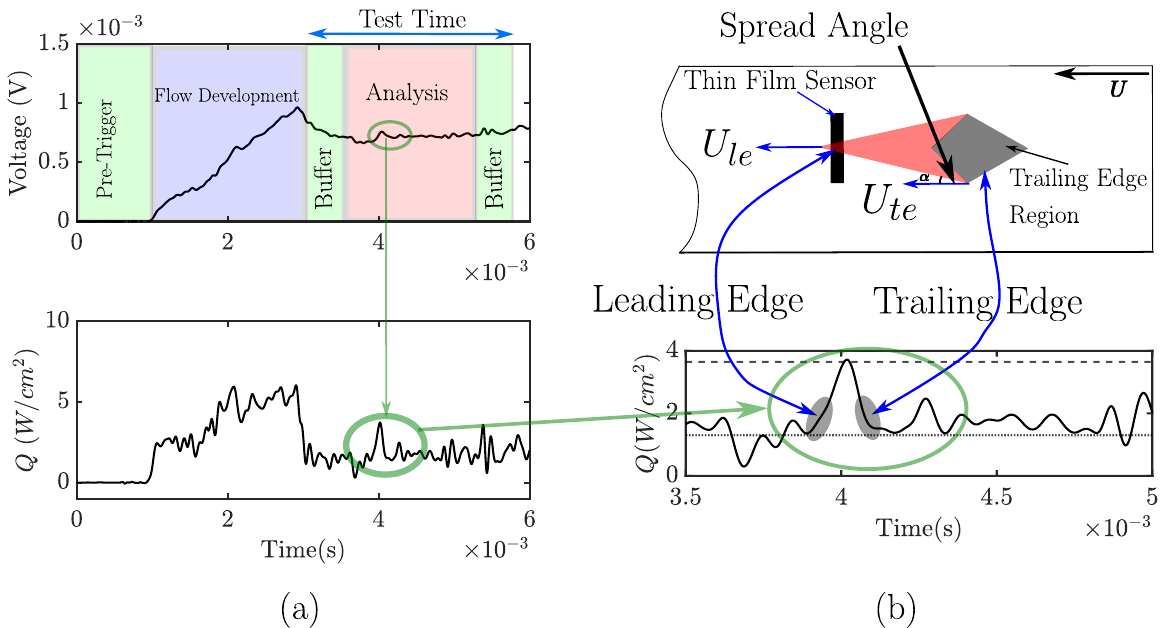}}
  \captionsetup{style=capcenter}
  \caption{(a) Typical raw signal and post-processed heat transfer time history obtained through a thin film sensor. (b) Zoomed in view of a turbulent spot identified from heat transfer time history and its association with various features of a typical turbulent spot. $U_{le}$ and $U_{te}$ are speeds assciated with the leading and trailing edge of the turbulent spots respectively. }
\label{fig:spot_charac}
\end{figure}

\begin{figure}[H]
  \centerline{\includegraphics[width=\textwidth]{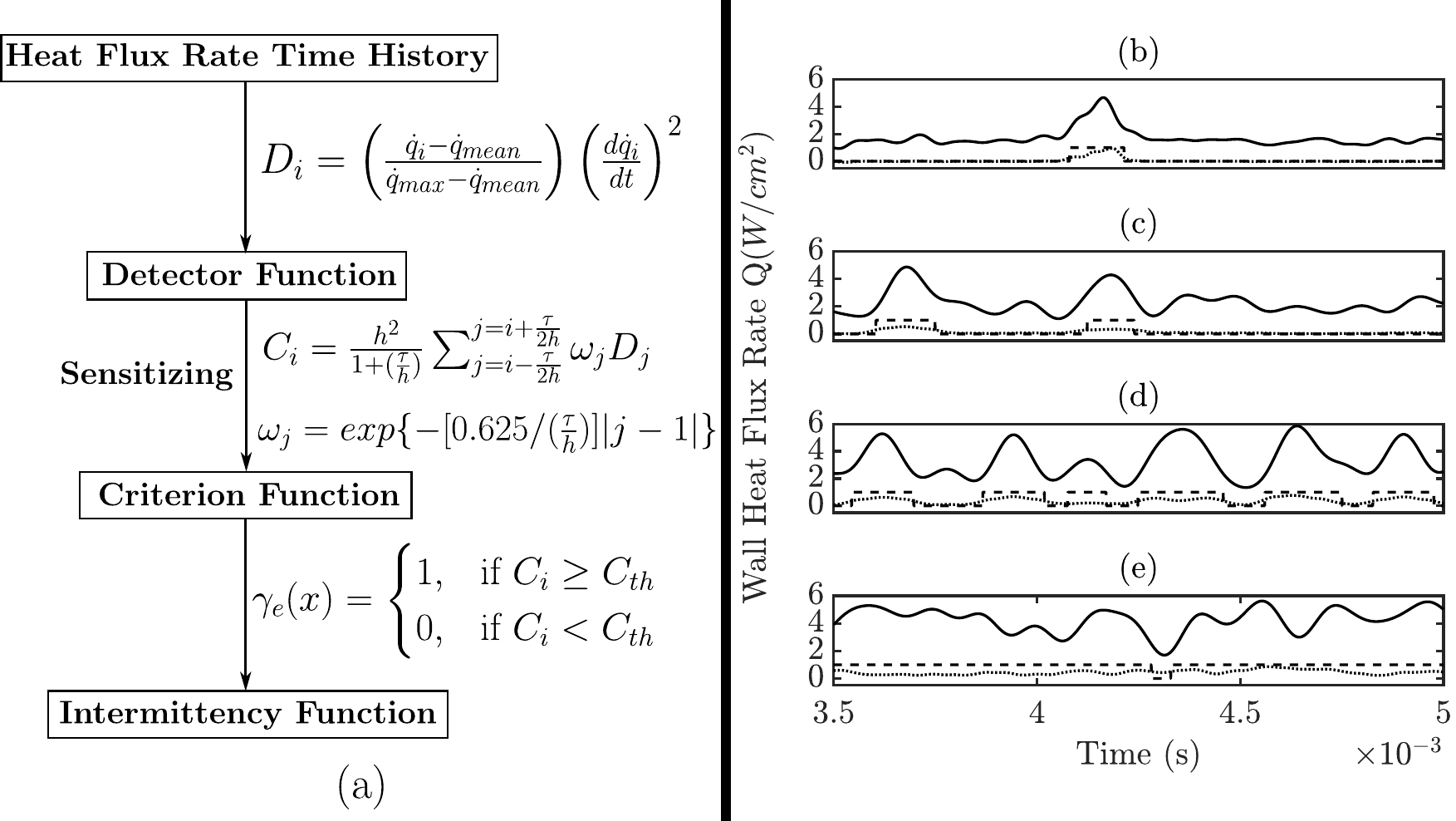}}
  \captionsetup{style=capcenter}
  \caption{(a) Work flow involved in Intermittency calculation from heat transfer time history. $D_{i}$ and $C_{i}$ are detector function and criterion function for $i^{th}$ data point respectively. $\gamma_{e}$ is intermittency calculated from experimental measurement of surface heat transfer. (b,c,d,e) Typical heat transfer time histories along with criterion function (\dotted) and intermittency (\dashed) time histories from sensors placed in transitional boundary layers developed over flat plate (d,e) and cone (b,c).}
\label{fig:inter_typ_ht}
\end{figure}

\begin{figure}[h]
  \centerline{\includegraphics[width=0.85\textwidth]{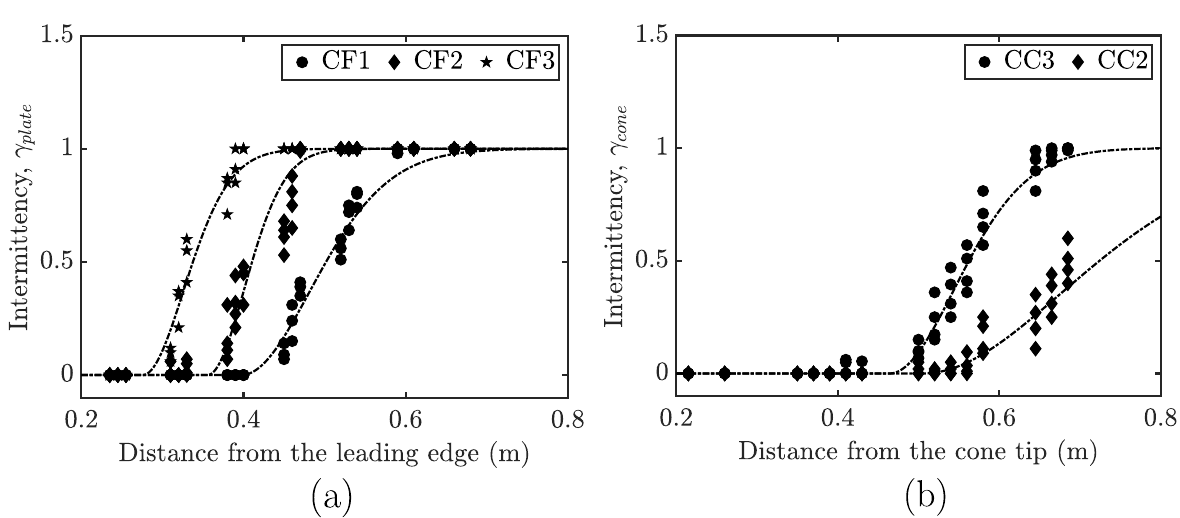}}
  \captionsetup{style=capcenter}
  \caption{Spatial distribution of intermittency along the length of (a) flat plate (b) cone for all the experimental runs. (\dashddot) line represents universal intermittency for one of the shots for every experimental condition.} 
\label{fig:inter_cone_fp_comp}
\end{figure}

\hyperref[fig:spot_charac]{Figure \ref{fig:spot_charac}} shows the typical heat transfer time history acquired from thin film sensors and its association with convection of a turbulent spot over the sensor location. The deviation of heat transfer from theoretical laminar levels marks the passage of the leading edge of the spot over the sensor location and the return of heat transfer to the theoretical laminar levels marks the passage of trailing edge of the spot.
\hyperref[fig:spot_charac]{Figure \ref{fig:spot_charac}} shows that the heat transfer values deviate from theoretical laminar level between $3.8$-$4.3$ ms due to passage of a turbulent spot over the sensor. Heat transfer time histories were used to quantify intermittency ($\gamma$), which is a measure of the fraction of boundary layer that has transitioned into fully turbulent state at a given location on a test model. The procedure for calculation of intermittency for both flat plate and cone is same and is outlined in \hyperref[fig:inter_typ_ht]{Figure \ref{fig:inter_typ_ht}}\textit{a}. The procedure closely follows the methodology outlined in Clark \textit{et al.} \citeyearpar{Clark94} except for the definition of detector function ($D_{i}$) which has been modified in the present work. The detector function in Clark \textit{et al.} \citeyearpar{Clark94} scales the time derivative of heat transfer with difference of theoretical laminar and turbulent heat transfer values usually evaluated with methods due to Eckert, van Driest or Spalding (Higgins \citeyear{ht_theory}). In the present work, the detector function expression scales the time derivative of heat transfer with the difference between local maxima and mean value of heat transfer. This eliminates the variance in intermittency calculation introduced due to difference in the methods of estimating theoretical heat transfer values. The temporal derivative of heat transfer was calculated using second order finite difference methods. From the detector function, a criterion function ($C_{i}$) was defined which is an exponential weighted ($\omega_{j}$) moving average based on sampling interval ($h$) and smoothing period ($\tau=20 \mu s$). The purpose of criterion function is to sensitize the detector function to efficiently distinguish between turbulent and non-turbulent regions of the boundary layer. The criterion function is normalized by its maximum value obtained from the data of all the sensors for a particular experiment. The intermittency is then set to $1$ if the criterion function crosses a certain threshold ($C_{th}$) otherwise it is set to $0$. In the present work, the value of $C_{th}$ is taken as $0.1$ times the maximum value of the normalized criterion function obtained in a particular experimental run. The intermittency is also set to $1.0$ if the heat transfer level at a particular sensor location crosses $0.8$ times the theoretical laminar heat transfer levels calculated from Eckert's method.  Typical result from the application of this work flow for calculation of intermittency time history from heat transfer time history is shown in \hyperref[fig:inter_typ_ht]{figure \ref{fig:inter_typ_ht}}\textit{b,c,d,e}.
 From \hyperref[fig:inter_typ_ht]{figure \ref{fig:inter_typ_ht}}\textit{b,c,d,e} it is clear that the methodology followed in the present work efficiently captures the intermittency associated with the transitional boundary layers encountered in the present work. Average of intermittency time history results in the intermittency at a sensor location. A spatial distribution of intermittency is thus obtained as shown in \hyperref[fig:inter_cone_fp_comp]{figure \ref{fig:inter_cone_fp_comp}}. In \hyperref[fig:inter_cone_fp_comp]{figure \ref{fig:inter_cone_fp_comp}}, the spatial distribution of intermittency obtained through all the experiments conducted in present campaign is compared with the universal intermittency distribution on the flat plate and cone. The universal intermittency distribution for flat plate ($\gamma_{plate}$) and cone ($\gamma_{cone}$) is calculated based on the following analytical relations
\begin{equation}
    \gamma_{plate} = 1.0 - exp\left[-(x-x_{tr})^{2}\frac{n\sigma}{U_{e}}\right]
    \label{eqn:inter_fp_analytical}
\end{equation}
\begin{equation}
    \gamma_{cone} = 1.0 - exp\left[-\frac{n\sigma}{U_{e}}x_{tr}\left(ln\frac{x}{x_{tr}}\right)(x-x_{tr})\right]
    \label{eqn:inter_cone_analytical}
\end{equation}

In \hyperref[eqn:inter_fp_analytical]{equations \ref{eqn:inter_fp_analytical}} and \ref{eqn:inter_cone_analytical}, $n$ refers to spot generation parameter, $U_{e}$ refers to boundary layer edge speed, $x_{tr}$ refers to transition onset location and $\sigma$ denotes spot growth parameter defined by \hyperref[eqn:sigma_def]{equation \ref{eqn:sigma_def}} (Vinod and Govindarajan \citeyear{sigma_def_ref})
\begin{equation}
    \sigma = \left[\frac{1}{U_{te}}-\frac{1}{U_{le}}\right]U_{e}tan{\alpha}
    \label{eqn:sigma_def}
\end{equation}
In \hyperref[eqn:sigma_def]{equation \ref{eqn:sigma_def}}, $\alpha$ is the spot spread angle defined as a function of boundary layer edge Mach number in \hyperref[eqn:spot_spread_angle]{equation \ref{eqn:spot_spread_angle}} (Doorly and Smith \citeyear{alpha_def_ref}). Experimental calculation of leading ($U_{le}$) and trailing ($U_{te}$) edge speed is discussed in the next section.
\begin{equation}
    \alpha = 3^{-3/2}2^{1/2}M_{e}^{-1}
    \label{eqn:spot_spread_angle}
\end{equation}

\subsection{Turbulent Spot Speeds, Streamwise Length Scales and Generation Rates}
\begin{figure}[h]
  \centerline{\includegraphics[width=0.85\textwidth]{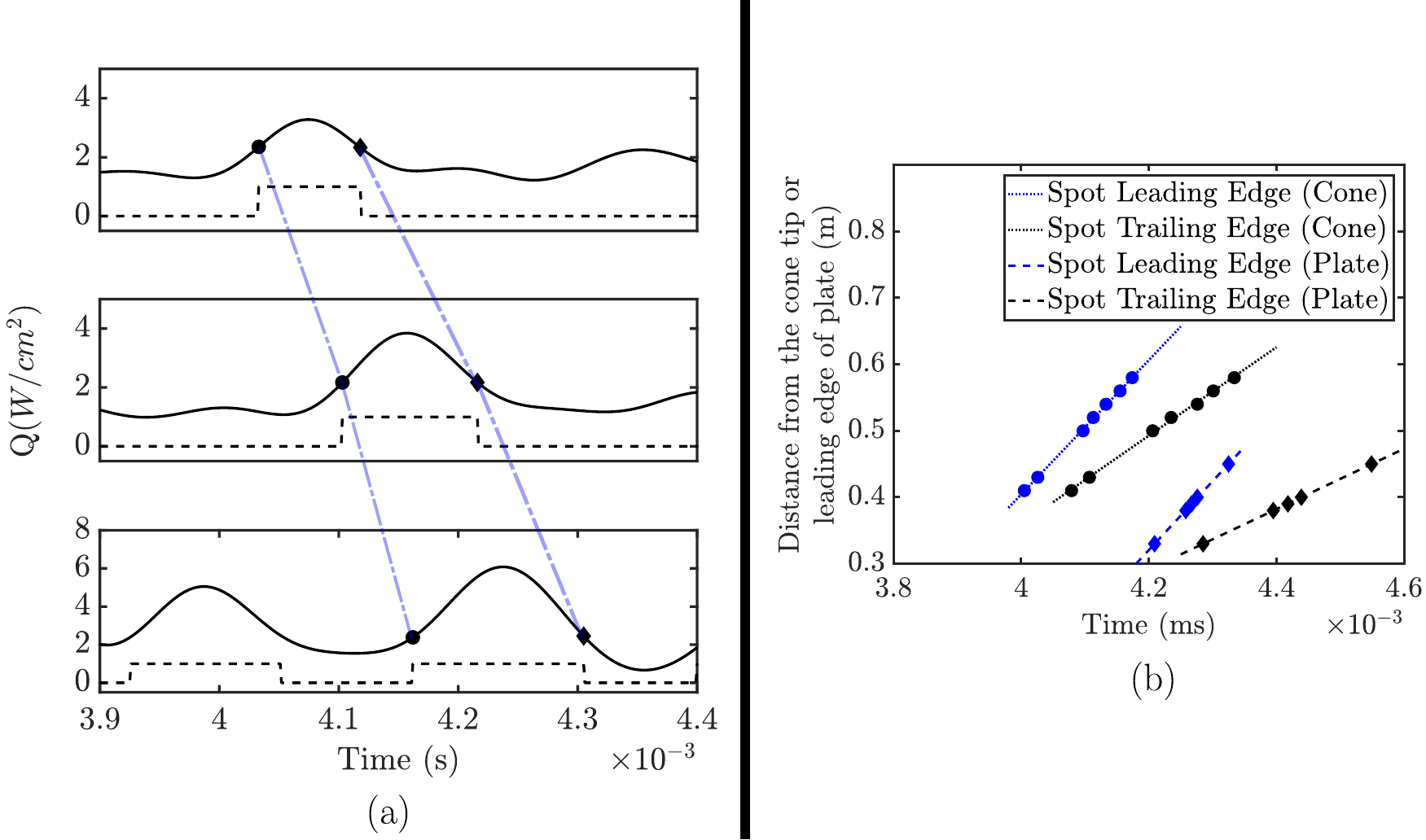}}
  \captionsetup{style=capcenter}
  \caption{(a) Heat transfer and intermittency time history capturing the convection of a turbulent spot over the cone. The first two sensors are 70 mm apart and the second and third sensors are 60 mm apart. Leading edge (\fullcirc) and trailing edge (\fulldiamond) of the spot is identified using the intermittency levels. Intermittency over each sensor location is shown by (\dashed). (b) Distance vs time trace of leading and trailing edge of a typical spot developed on cone and flat plate. The leading and trailing edge speeds are calculated from the slope of these traces shown by (\dotted) for cone and (\dashed) for flat plate} 
\label{fig:turb_spot_convect_cone}
\end{figure}

\begin{figure}[H]
  \centerline{\includegraphics[width=0.85\textwidth]{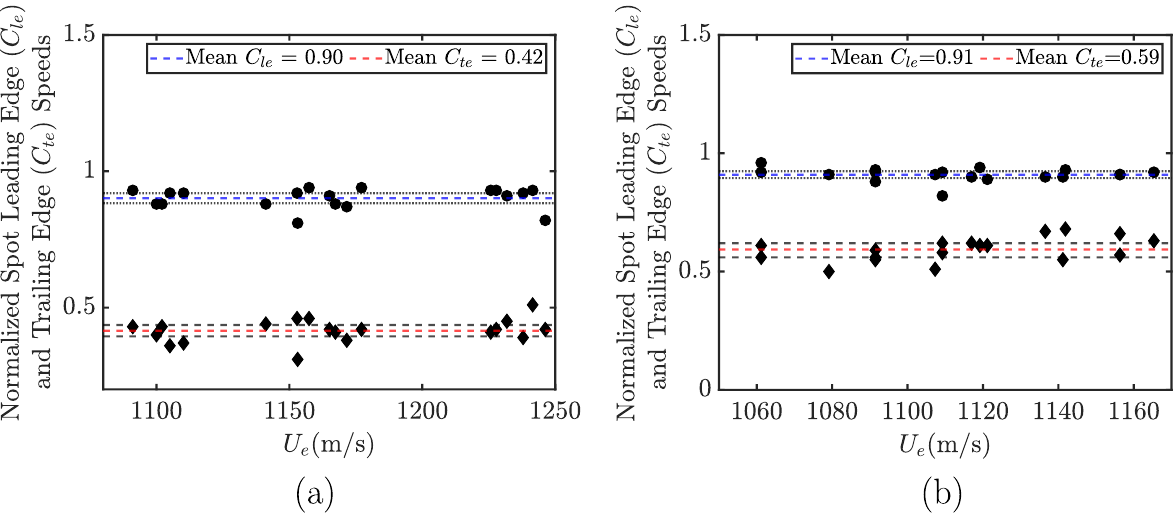}}
  \captionsetup{style=capcenter}
  \caption{Normalized spot leading edge (\fullcirc) and trailing edge (\fulldiamond) speeds of turbulent spots identified in all the experiments conducted over (a) flat plate and (b) cone. In both the figures (\dotted) and (\dashed) shows the lower and upper bounds of 95\% confidence interval for normalized leading edge and trailing edge speeds respectively.} 
\label{fig:all_spot_cone_fp}
\end{figure}

\hyperref[fig:turb_spot_convect_cone]{Figure \ref{fig:turb_spot_convect_cone}}\textit{a} shows typical convection of a turbulent spot identified by using heat transfer and intermittency time histories of three heat transfer sensors placed over the surface of cone. The leading edge and trailing edge of the turbulent spot is identified by the intermittency switch. Similar technique was used in case of flat plate as well. The distance between the sensors and the time taken by leading and trailing edge of the spot to traverse the distance between the sensors is known. The speeds associated with the leading edge ($U_{le}$) and trailing edge ($U_{te}$) of the turbulent spots can be calculated. In the present work, 19 distinct turbulent spots were identified in all the experiments for the flat plate and 16 distinct spots were identified for cone. For each one of the spots, a distance time trace for leading and trailing edge of the spot was plotted as shown in \hyperref[fig:turb_spot_convect_cone]{figure \ref{fig:turb_spot_convect_cone}}\textit{b}, the slope of which is the corresponding speeds of the spot. The distance time trace, for each turbulent spot, was plotted for its convection across 2-5 heat transfer sensors. The calculated leading and trailing edge speeds were normalized with boundary layer edge speed to result in the normalized leading ($C_{le}$) and trailing ($C_{te}$) edge speeds associated with turbulent spots for both flat plate and cone as shown in \hyperref[fig:all_spot_cone_fp]{Figure \ref{fig:all_spot_cone_fp}}.

From \hyperref[fig:all_spot_cone_fp]{figure \ref{fig:all_spot_cone_fp}} it is evident that for a given boundary layer edge speed, the leading edge of the turbulent spots developed on the cone and flat plate are convected at comparable speeds, while the trailing edge of the spots on flat plate moves with a slower speed than those developed on the cone. It is evident, from the spot velocity data, that the spots developed in a planar boundary layer have a larger streamwise growth rates than their axisymmetric counterpart. A first order approximation of the length of the turbulent spots can further elucidate this aspect. The maximum and minimum streamwise length scales ($l_{max}$ and $l_{min}$) associated with the turbulent spots can be estimated by \hyperref[eqn:lmax]{equations \ref{eqn:lmax}} and \ref{eqn:lmin} (Fiala \textit{et al.}\citeyear{Fiala06}). In \hyperref[eqn:lmax]{equations \ref{eqn:lmax}} and \ref{eqn:lmin} and subsequent equations, $t_{te}$ and $t_{le}$ are the time stamps where the trailing edge and the leading edge of the spot crosses the sensor location.
\begin{equation}
    l_{max}=U_{le}(t_{te}-t_{le})    
    \label{eqn:lmax}
\end{equation}
\begin{equation}
    l_{min}=U_{te}(t_{te}-t_{le})    
    \label{eqn:lmin}
\end{equation}
The convection speed of a turbulent spot varies along its length and in the present work it is assumed that the spot convects at a speed ($U_{sc}$) which is mean of the leading edge and trailing edge speeds (\hyperref[eqn:scvel]{equation \ref{eqn:scvel}}). This assumption leads to a first order approximation of spot streamwise length scales, given by \hyperref[eqn:tslen]{equation \ref{eqn:tslen}}, which is reasonable enough to compare streamwise growth rate of turbulent spots generated in flat plate and cone. 
\begin{equation}
    U_{sc}=0.5(U_{le}+U_{te})
    \label{eqn:scvel}
\end{equation}
\begin{equation}
    l_{ts} \approx U_{sc}(t_{te}-t_{le})
    \label{eqn:tslen}
\end{equation}

\begin{figure}[H]
  \centerline{\includegraphics[width=0.85\textwidth]{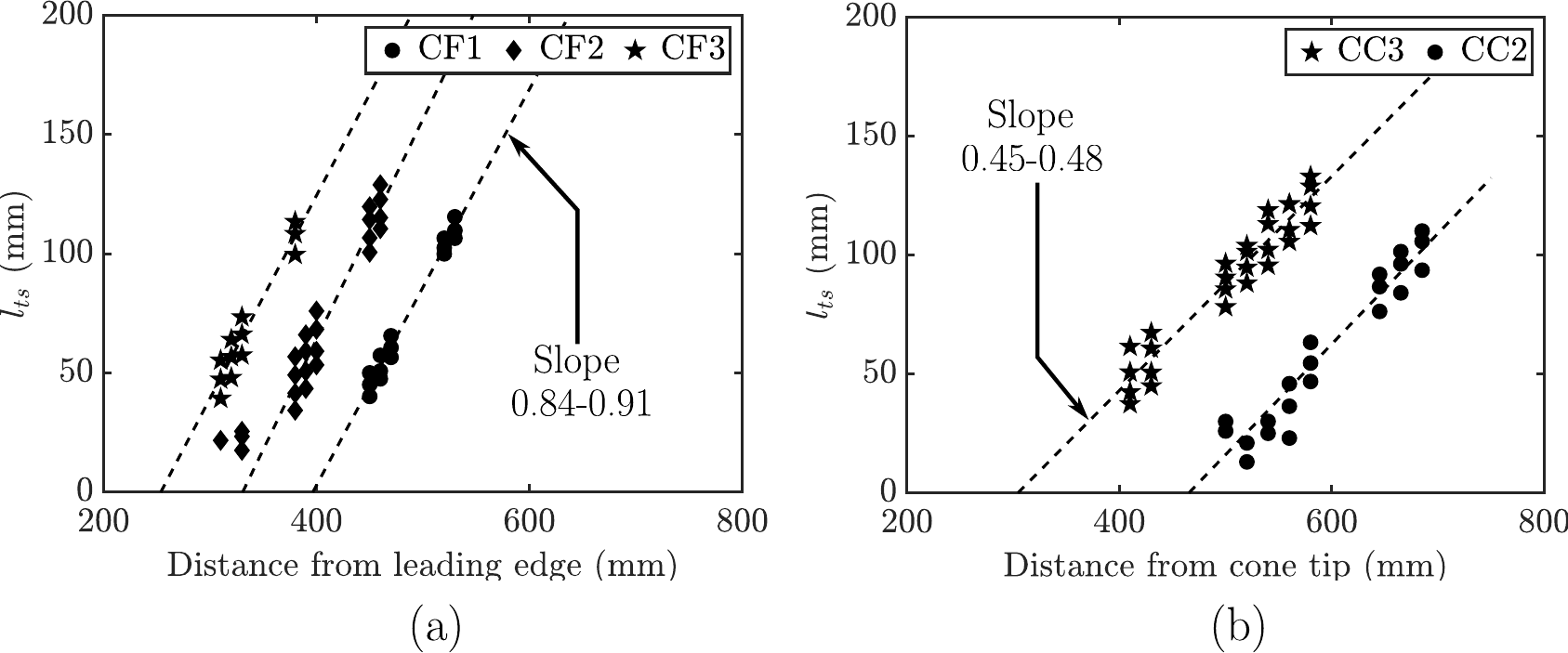}}
  \captionsetup{style=capcenter}
  \caption{Streamwise length scale estimates of turbulent spots identified from \hyperref[eqn:tslen]{equation \ref{eqn:tslen}} at various sensor locations along the (a) flat plate surface (b) cone surface. The (\dashed) line shows the linear fit to the spot length data with slope ranges indicated in the figure. The intersection of (\dashed) line with x-axis shows the location of generation of the turbulent spots on particular test model surface.} 
\label{fig:spot_len}
\end{figure}

\begin{figure}[H]
  \centerline{\includegraphics[width=\textwidth]{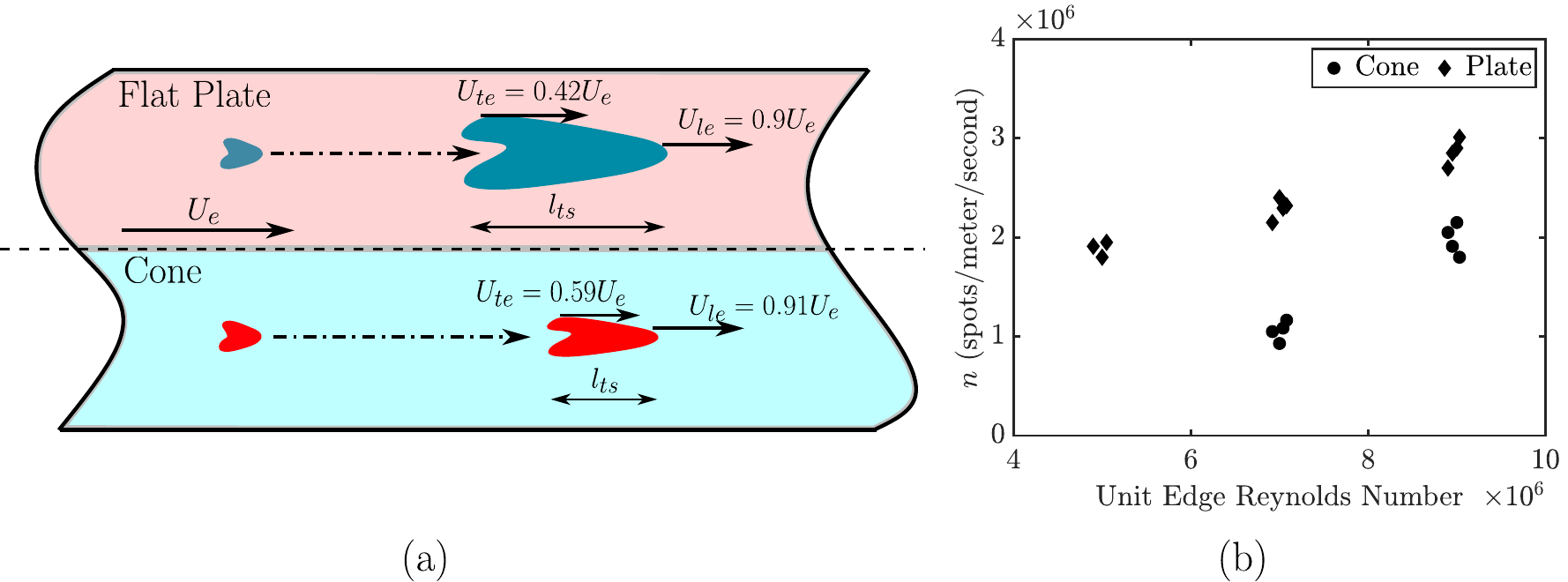}}
  \captionsetup{style=capcenter}
  \caption{(a) Schematic depicting growth of a typical turbulent spot developed on a flat plate and cone surface over the same longitudinal length. (b) Turbulent spot generation rate for flat plate (\fulldiamond) and cone (\fullcirc).} 
\label{fig:concl_fig}
\end{figure}

The streamwise length estimates obtained from \hyperref[eqn:tslen]{equation \ref{eqn:tslen}} at various locations on flat plate and cone are shown in \hyperref[fig:spot_len]{figure \ref{fig:spot_len}}. A quick comparison between \hyperref[fig:spot_len]{figure \ref{fig:spot_len}}\textit{a} and \hyperref[fig:spot_len]{figure \ref{fig:spot_len}}\textit{b} shows that turbulent spots generated on a flat plate experience a larger streamwise growth rate along the length of the plate owing to their slower trailing edge speed when compared to the spots generated on the cone. This aspect is clearly elucidated in the schematic shown in \hyperref[fig:concl_fig]{figure \ref{fig:concl_fig}}\textit{a}. Here the turbulent spot developed on a flat plate enlarges more when compared to the spot developed on the cone if both the spots traverse a similar length along the plate and cone respectively. It is to be noted that the transitional boundary layer location and length is different for the plate and cone.\\
The spot generation rates in the present work have been calculated from \hyperref[eqn:inter_fp_analytical]{equations \ref{eqn:inter_fp_analytical}} and \ref{eqn:inter_cone_analytical}. The RHS of both \hyperref[eqn:inter_fp_analytical]{equations \ref{eqn:inter_fp_analytical}} and \ref{eqn:inter_cone_analytical} is determined from experimental data with the procedure outlined in \hyperref[sec:ht_inter_calc]{section \ref{sec:ht_inter_calc}}. All the quantities in the RHS of both \hyperref[eqn:inter_fp_analytical]{equations \ref{eqn:inter_fp_analytical}} and \ref{eqn:inter_cone_analytical} have been calculated from present experiments except for the spot generation rate parameter, $n$. Hence the value of spot generation parameter can be calculated by using the experimental outcomes of the present work and the same is shown in \hyperref[fig:concl_fig]{figure \ref{fig:concl_fig}}\textit{b}. For the same unit boundary layer edge Reynolds number, higher number of turbulent spots are generated on a flat plate in comparison to cone. It is also evident from \hyperref[fig:concl_fig]{figure \ref{fig:concl_fig}}\textit{b} that with increasing unit edge Reynolds number the number of turbulent spots generated in the planar and axisymmetric boundary layer approach the same value. This indicates that at certain high enough edge unit Reynolds the transitional boundary layer length in both the test models will be entirely governed by the streamwise growth of the turbulent spots.\\
Earlier studies have shown that the transition onset Reynolds number for flat plate is lower than the one encountered in the case of cone. Stability computations have also revealed that planar boundary layers are more receptive to the freestream noise in comparison to the axisymmetric conical boundary layers (Chen \textit{et al.} \citeyear{fp_cone_comp}). Hence earlier transition onset is witnessed in planar boundary layer in comparison to axisymmetric boundary layer. A similar trend was found in the present study as well (\hyperref[fig:inter_cone_fp_comp]{figure \ref{fig:inter_cone_fp_comp}}). Furthermore, in the present study the transitional boundary layer length over flat plate was found to be shorter than its axisymmetric counterpart. The spot generation rate results reported over here in combination of spot growth rate aspects provide an explanation for transitional boundary layer length to be shorter in flat plate when compared to the cone. For the same edge Reynolds number, larger number of turbulent spots are generated on the flat plate in comparison to the cone (\hyperref[fig:concl_fig]{figure \ref{fig:concl_fig}}\textit{b}). After convecting through the same length a larger area would be occupied by the turbulent spots developed on the plate, owing to its higher longitudinal growth rate when compared to spots developed on the cone. The merging of turbulent spots to give rise to a fully developed turbulent boundary layer occurs over a shorter distance in flat plate when compared to the cone owing to the aforementioned outcomes and observations. The experiments and analysis presented here provide a reason and explanation for the observations on boundary layer transition over flat plate and cone reported in earlier and present work. 

\section{Conclusions}
The present work offers a unique comparison between transitional planar and axisymmetric boundary layers in terms of intermittency and turbulent spot characteristics like spot leading and trailing edge speeds, growth of streamwise length scales and spot generation rates. Surface heat transfer measurements were carried out using platinum thin film sensors to ascertain the nature of the boundary layer developed on the test models. A methodology to calculate the intermittency from surface heat transfer measurements was outlined and a modified detector function definition was proposed and used in the present work. It was found that the modified detector function efficiently captured intermittency generated by the convection of turbulent spots over the heat transfer sensor location. Spatial variation of intermittency obtained from the experiments was compared with universal intermittency spatial distribution and a satisfactory match was found for both the test models. Transition onset locations on both flat plate and cone for all the experimental conditions was identified by using the intermittency measurements. The transition onset on plate boundary layer occurred earlier than its axisymmetric counterpart. Furthermore, it was found that the planar transitional boundary layer was shorter than the axisymmetric transitional boundary layer. Characteristics of turbulent spots, like their leading edge and trailing edge speeds, length scales and spot generation rates, were calculated for both the test modes. The leading edge of the turbulent spots developed on both flat plate and cone convected at similar speeds which, in the present work, was found to be about 90\% of the boundary layer edge speed. It was found that the trailing edge of the spot developed in a planar boundary layer convected at a lower speed than the spot developed on an axisymmetric boundary layer. This difference in trailing edge convection speed of turbulent spots affects their streamwise growth rate with a higher growth rate for spots developed over flat plate. It was also found that for the similar boundary layer edge Reynolds number and Mach number, higher number of turbulent spots are generated on planar boundary layers in comparison to axisymmetric boundary layers. These outcomes of present work provide an explanation for shorter transition lengths observed in flat plate when compared to the cone. 

\begin{bmhead}[Acknowledgments]
The authors acknowledge funding and support from DRDL, Hyderabad to carry out this work.
\end{bmhead}

\begin{bmhead}[Declaration of Interests]
The authors declare no conflict of interests.
\end{bmhead}



\end{document}